\documentclass{aa}
\usepackage{graphicx}

\begin{document}

\title{Element abundances in cool white dwarfs}

\subtitle{II. Ultraviolet observations of DZ white dwarfs\thanks{
Based partly on observations made with the NASA/ESA Hubble Space Telescope,
obtained from the data archive at the Space Telescope Science Institute.
StScI is operated by the Association of Universities for Research in
Astronomy, Inc., under NASA contract NAS\,5-26555.
DK was Visiting Astronomer at the German-Spanish Astronomical Centre,
Calar Alto, operated by the Max-Planck-Institute for Astronomy, Heidelberg,
jointly with the Spanish National Commission for Astronomy.}}

\author{Burkhard Wolff\inst{1}
        \and
        Detlev Koester\inst{1}
	\and
	James Liebert\inst{2}
       }

\offprints{B. Wolff}

\institute{Institut f\"ur Theoretische Physik und Astrophysik,
           Universit\"at Kiel, D-24098 Kiel, Germany\\
           \email{wolff, koester@astrophysik.uni-kiel.de}
      \and Steward Observatory, University of Arizona,
           933 North Cherry Avenue, Tucson, AZ 85721, USA\\
	   \email{liebert@as.arizona.edu}
          }

\date{Received 21 December 2001 / Accepted 31 January 2002}

\abstract{We present a small data base of homogeneously derived photospheric
element abundances of DZ white dwarfs and related objects. Our previous
investigations are supplemented with the analysis of
ultraviolet spectra for nine white dwarfs.
Of particular interest is the detection of L$\alpha$ absorption in
van~Maanen~2 and a determination of the effective temperature of this star.
The new value is about 1000~K lower than previous results due to the strong
ultraviolet absorption by metals which has to be considered consistently. 
The metal abundances of our sample stars are compatible with the predictions
from the two-phase accretion model of
Dupuis et al. (\cite{Dupuis92}, \cite{Dupuis93a}, \cite{Dupuis93b}).
Small deviations can be observed for the abundance ratios in some objects.
This could indicate non-solar metal-to-metal ratios in the accreted material.
Hydrogen can be detected in virtually all of our objects. However,
its average accretion rate must be at least two orders of magnitude lower
than the metal accretion rate.
\keywords{stars: abundances -- stars: atmospheres -- white dwarfs --
          ultraviolet: stars}
         }

\maketitle

\sloppy

\section{Introduction}

At the cool end of the white dwarf sequence traces of heavier elements can be
found in several objects with hydrogen- and helium-rich atmospheres.
The presence of metals is usually attributed to
accretion from denser parts of
the interstellar medium combined with diffusion into deeper layers.
This idea has been put on a sound theoretical basis by
Dupuis et al. (\cite{Dupuis92}, \cite{Dupuis93a}, \cite{Dupuis93b})
whose two-phase accretion model assumes that
accretion of metals is low 
($\dot{M}\approx 5\cdot10^{-20}\,M_{\odot}\,{\rm yr}^{-1}$) most of the time
and is larger
($\dot{M}\approx 5\cdot10^{-15}\,M_{\odot}\,{\rm yr}^{-1}$)
only during rare and short passages through interstellar clouds.
Since the accreted elements diffuse downwards on
time scales of typically $10^6$ years they are only visible during or
shortly after such encounters.

Empirical tests of the accretion model benefit from
different diffusion time scales of the elements so that photospheric
abundances and abundance ratios vary in a characteristic way with time
after the encounter.
Dupuis et al. (\cite{Dupuis93b}) concluded that -- within the large
observational uncertainties -- the two-phase accretion model can account
for the observed abundances and their ratios.
The most notable remaining problem for the helium-rich objects (spectral types
DZ and DBZ) is the accretion of hydrogen.
As the lightest element, hydrogen is not subject to downward diffusion and
should be accumulated in the upper atmospheric layers. However, hydrogen is only
rarely observed in DZ stars and the measured abundances are usually
rather low.

A general problem of previous comparisons is the uncertainty arising from
observations with low signal-to-noise ratios and model atmospheres not
including recent advances in input physics. In Paper\,I
(Koester \& Wolff \cite{KW}),
we have concluded that it is worthwhile to study the abundance patterns in
DZ white dwarfs again using new or newly calibrated observations and
the latest improvements in model atmosphere calculations. The ultimate goal
is a homogeneous data base of element abundances for the comparison with
the two-phase accretion model.

\begin{table*}[t]
\caption[]{Observations}
\begin{flushleft}
\begin{tabular}{l@{}c@{}llllll}
\noalign{\smallskip}
\hline
\noalign{\smallskip}
\multicolumn{3}{l}{WD Number} &Name &Observations &$\lambda / $\,\AA\ 
                              &$\Delta \lambda / $\,\AA\ &Remarks\\
\noalign{\smallskip}
\hline
\noalign{\smallskip}
0002 &+   &729 &GD\,408    &IUE, SWP19006L     &1150--1950  &6.0 \\
     &    &    &           &IUE, LWR15058L     &1850--3250  &6.0 \\
     &    &    &           &MMT                &3750--4650  &1.0
                           &Sion et al. (\cite{Sion88}) \\
0038 &$-$ &226 &LHS\,1126  &HST/FOS, G190H     &1573--2330  &1.5 \\
     &    &    &           &HST/FOS, G270H     &2221--3301  &2.1 \\
     &    &    &           &HST/FOS, G400H     &3240--4822  &3.1 \\
     &    &    &           &BVRIJHK photometry &4400--22000 &
                           &Bergeron et al. (\cite{Bergeron97}) \\
0046 &+   &051 &vMa\,2     &IUE, LWR03322L     &1850--3250  &6.0 \\
     &    &    &           &IUE, LWR06474L     &1850--3250  &6.0 \\
     &    &    &           &Calar Alto, 3.5m   &3730--5075  &4.0 \\
     &    &    &           &BVRIJHK photometry &4400--22000 &
                           &Bergeron et al. (\cite{Bergeron97}) \\
0552 &$-$ &041 &LP\,658-2  &HST/FOS, G190H     &1573--2330  &1.5 \\
     &    &    &           &HST/FOS, G270H     &2221--3301  &2.1 \\
     &    &    &           &HST/FOS, G400H     &3240--4822  &3.1 \\
     &    &    &           &BVRIJHK photometry &4400--22000 &
                           &Bergeron et al. (\cite{Bergeron97}) \\
0752 &$-$ &676 &BPM\,4729  &HST/FOS, G190H     &1573--2330  &1.5 \\
     &    &    &           &HST/FOS, G270H     &2221--3301  &2.1 \\
     &    &    &           &HST/FOS, G400H     &3240--4822  &3.1 \\
     &    &    &           &BVRIJHK photometry &4400--22000 &
                           &Bergeron et al. (\cite{Bergeron97}) \\
1011 &+   &570 &GD\,303    &IUE, SWP18994L     &1150--1950  &6.0 \\
     &    &    &           &IUE, LWR15047L     &1850--3250  &6.0 \\
     &    &    &           &MMT                &3750--4650  &1.0
                           &Sion et al. (\cite{Sion88}) \\
1225 &$-$ &079 &K\,789-37  &IUE, SWP30001L     &1150--1950  &6.0 \\
     &    &    &           &IUE, LWP09836L     &1850--3250  &6.0 \\
     &    &    &           &SAAO, 1.9m         &3600--5000  &5.0 
                           &Koester et al. (\cite{Koester90}) \\
1822 &+   &410 &GD\,378    &IUE, SWP19549L     &1150--1950  &6.0 \\
     &    &    &           &IUE, LWR15578L     &1850--3250  &6.0 \\
     &    &    &           &MMT                &3750--4650  &1.0
                           &Sion et al. (\cite{Sion88}) \\
2216 &$-$ &657 &L\,119-34  &IUE, SWP29720L     &1150--1950  &6.0 \\
     &    &    &           &IUE, LWR14743L     &1850--3250  &6.0 \\
\noalign{\smallskip}
\hline
\noalign{\smallskip}
\label{tabobs}
\end{tabular}
\end{flushleft}
\end{table*}

We started in Paper\,I with the Hubble Space Telescope (HST) observations
of the DZA white dwarfs L\,745-46A and Ross\,640. Both objects show
weak Balmer lines in addition to metals
and are therefore interesting with regard to
the hydrogen problem mentioned above. We have used the high quality HST spectra
to test our analysis methods and to study the remaining uncertainties of
the model atmospheres. In comparison with previous investigations, we were
able to determine more accurate element abundances, especially for carbon and
iron. Major improvements were possible for the modeling of the broad
L$\alpha$ lines in both stars which could be reproduced for the first time
without any unphysical assumptions. This provides new means for the analysis
of the hydrogen content in cool white dwarfs.

In this paper, we analyze all available ultraviolet observations of white
dwarfs with spectral types DZ and DBZ. The ultraviolet is very suited for
the study of metal abundances because the relevant elements have in general
much stronger lines in this part of the electromagnetic spectrum than in the
optical part -- with the exception of the Ca\,II H and K lines.
We use for most stars observations with the
International Ultraviolet Explorer (IUE) which have been newly calibrated
(NEWSIPS) and which show significant improvements compared to older
versions. HST/FOS observations could also be used for three stars.

\section{Model atmospheres}
\label{secatm}

A general description of the procedures used for the calculation of our
LTE model atmospheres and synthetic spectra has been given by
Finley et al. (\cite{FKB}). Further details can be found in
Paper\,I. We only explain briefly the effects of different equations of
states (EOS) and the treatment of L$\alpha$ absorption.

\begin{figure*}[tbp]
\centering
\includegraphics[width=\textwidth]{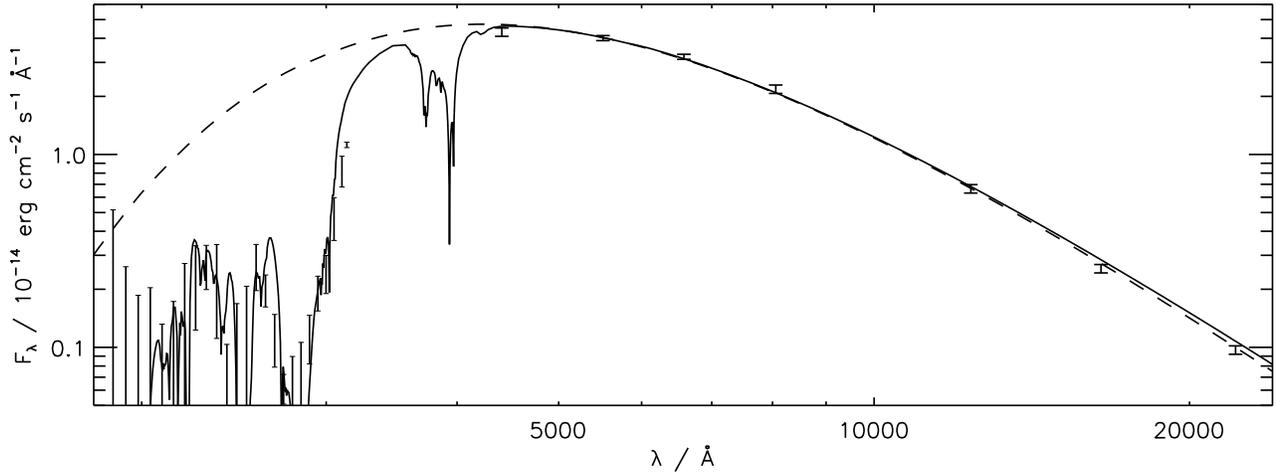}
\caption{Flux distribution of van~Maanen~2: $BVRIJHK$ photometry from
Bergeron et al. (\cite{Bergeron97}) and IUE observations (rebinned to
60~\AA\ resolution). The continuous curve is a model with $T_{\rm eff}=5700$~K,
metals, and hydrogen, the dashed curve a pure helium model with
$T_{\rm eff}=6700$~K}
\label{figvmflux}
\end{figure*}

We use for our models a version of the Hummer-Mihalas-D\"appen
EOS (Mihalas et al. \cite{Mihalas90}) which predicts
larger non-ideal effects and larger ionization than the EOS by
Saumon \& Chabrier (\cite{SC94}) and Saumon et al. (\cite{SCVH95}).
In Paper\,I, we have compared the results of our standard models with
models were non-ideal effects are switched off and the ionization fraction is
determined only by thermal effects using the unmodified Saha equation.
For models with effective temperatures of about 8000~K non-ideal effects are
negligible but the situation may be different at the lower temperatures
of some of the stars in this paper. We therefore test the influence of
non-ideal effects in these cases.

Of importance for the analysis of the HST spectra of L\,745-46A and Ross\,640
has been the L$\alpha$ wing which dominates the UV spectra of both 
stars. Since model calculations using van-der-Waals broadening by neutral helium
result in
a L$\alpha$ wing extending far into the optical region we have developed
in Paper\,I a new approach to model this line. Since we are only interested
in the far wings we use the quasistatic limit of the broadening theory
(see e.g. Allard \& Kielkopf \cite{AKi91}) but take into account two effects
which are important at high perturber densities: the variation of the
transition probability with distance between emitter and perturber and
the strongly repulsive potential at close distances. With this procedure
we were able to reproduce the L$\alpha$ wings in both stars at the same
abundances as derived from optical Balmer lines.

\section{Observations}

Three of our objects have been observed with the
Faint Object Spectrograph (FOS) onboard HST. For the remaining six
DZ/DBZ white dwarfs we have used low dispersion IUE spectra.
The IUE data were retrieved from the IUE final archive and have been calibrated
with the NEWSIPS calibration (Nichols et al. \cite{Nichols94}).
In addition to these
data, we have also used optical observations and optical photometric scans.
A list of the observations can be found in Table~\ref{tabobs}.

\section{Analysis}

\subsection{Van~Maanen~2}

Van~Maanen~2 (WD\,0046+051) is one of the three classic white dwarfs.
It is a cool and very metal-rich DZ star, the optical spectrum
exhibiting, besides the typical Ca\,II H+K lines, Ca\,I at 4227~\AA,
Fe\,I at 4384~\AA, and several Fe\,I lines on the blue wing of Ca\,II.
Analysis of this star, using model atmospheres, has been pioneered by
Weidemann (\cite{Weidemann60}) and later by Wegner (\cite{Wegner72}) and
Grenfell (\cite{Grenfell74}).
The ultraviolet spectrum is dominated by the strong Mg\,I line at
2852\,\AA. Also important are the Mg\,II lines at 2796/2803~\AA\ and several
weak lines of Fe\,I and Fe\,II. Until now, the ultraviolet spectrum
has not been analyzed in detail.
Bergeron et al. (\cite{Bergeron97}, \cite{Bergeron01}) have recently
determined $T_{\rm eff}=6770$~K
from $BVRIJHK$ photometry using a pure helium atmosphere. Combining this
result with the
trigonometric parallax they also derived $\log g = 8.40$.

We use the photometry together with the co-added IUE LWR spectra
to define the flux distribution of van~Maanen~2. An optical spectrum
(3730--5075~\AA) obtained at the Calar Alto observatory complements the
observations for the determination of metal abundances.
We started the analysis with a pure helium atmosphere of
$T_{\rm eff}=6700$~K. Fig.~\ref{figvmflux} shows the fit of this model to
the flux distribution. To account for the solid angle of the star
the model has been scaled to reproduce the
infrared J magnitude. It fits the photometric data well but
fails to reproduce the overall flux level at ultraviolet wavelengths where
the absorption from metals is important.
Then we calculated model atmospheres and spectra with
$T_{\rm eff}=6700$~K and
several chemical compositions to determine the abundances of magnesium, calcium,
and iron from optical and ultraviolet lines by a visual comparison.
It turned out that the backwarming effect from the ultraviolet absorption --
especially from the strong Mg\,I line -- and the effect of increased electron
density significantly alter the
flux distribution. The effective temperature has to be reduced and,
consequently, the abundances have to be determined again. In an iterative
process we finally derived $T_{\rm eff} = 5700$~K,
about 1000~K lower than the original result from Bergeron et al.

\begin{figure*}[tbp]
\centering
\includegraphics[width=\textwidth]{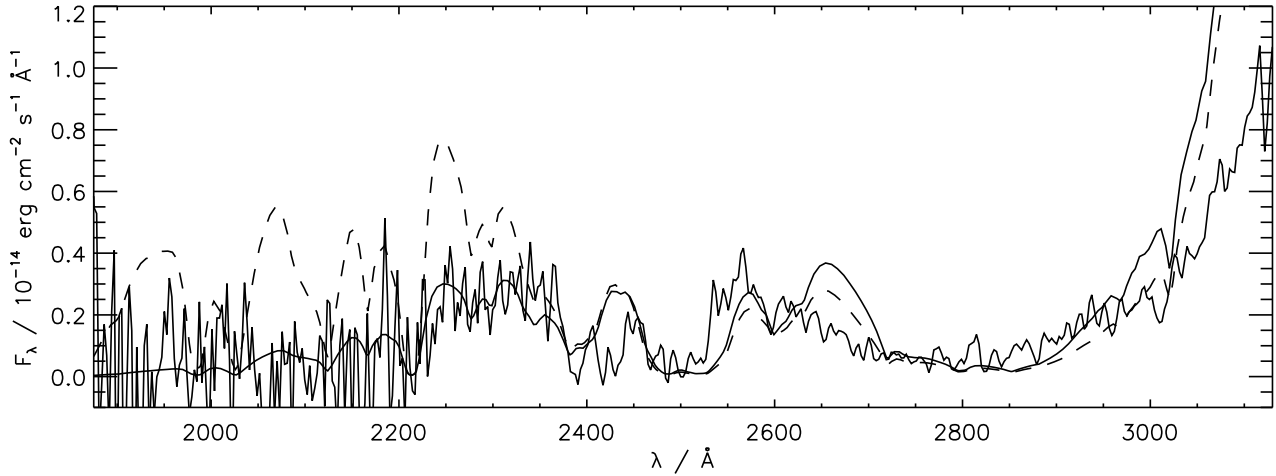}
\caption{IUE spectrum of van~Maanen~2 compared to a model with
$T_{\rm eff} = 5700$~K, $\log g = 7.9$, $\log {\rm H/He} = -5.0$,
$\log {\rm Mg/He} = -9.1$, $\log {\rm Si/He} = -9.0$, $\log {\rm Ca/He}=-10.65$,
$\log {\rm Fe/He} = -9.4$ (solid line) and a model with the same parameters
but without hydrogen (dashed line)}
\label{figvmuv}
\end{figure*}

Fig.~\ref{figvmflux} shows that the new model can account for the complete
observed flux distribution. At optical and infrared wavelengths both model
spectra are virtually identical.
Our effective temperature is similar to previous results from e.g.
Wegner (\cite{Wegner72}), Grenfell (\cite{Grenfell74}), and Liebert et al.
(\cite{Liebert87}) who considered the contributions of metals as electron
donators in their calculations. This shows the importance of an accurate
chemical composition for determining effective temperatures of
DZ white dwarfs and related objects -- a result which is also
found and discussed in more detail by Provencal et al. (\cite{Provencal})
in their analysis of Procyon\,B.

From a visual comparison of the fits to the flux distribution with different
models we estimate the error of $T_{\rm eff}$ to about $\pm$\,200~K. 
The solid angle necessary to reproduce the photometry
together with the trigonometric parallax from Hipparcos
($\pi = 226.95 \pm 5.35$~mas; ESA \cite{ESA}) and the mass-radius relation
from Wood (\cite{Wood94}; without hydrogen) implies a surface gravity of
$\log g = 7.9$. The uncertainties in $T_{\rm eff}$ and $\pi$ translate
into $\pm$\,0.2 for the log of the gravity.
The higher value of $\log g = 8.4$ derived
by Bergeron et al. (\cite{Bergeron97}, \cite{Bergeron01})
is due to the higher effective
temperature which gives a higher intrinsic flux and therefore a smaller
radius and a larger mass for the star. 

The absorption from magnesium, calcium, and iron alone cannot reproduce
the ultraviolet flux completely. As illustrated in Fig.~\ref{figvmuv},
there is a significant discrepancy at $\lambda \la 2300$~\AA. We have included
in the model calculations all lines from the identified metals and have also
determined an upper limit for silicon. One may speculate that other
unidentified metals may be responsible for the absorption. However, we do not
think this to be plausible since, for example, the absorption from C\,I at
1930~\AA\ would result in a competely different shape and other metals
than the ones considered here have not been observed so far in DZ white
dwarfs. With regard to the experience with absorption from L$\alpha$
in L\,745-46A and Ross\,640 (Paper\,I) we would rather attribute the missing
opacity to hydrogen. Fig.~\ref{figvmuv} shows that adding an abundance of
$\log {\rm H/He} = -5.0$ improves the fit considerably. The Balmer lines
are too weak at this abundance to be detected in existing optical spectra
so that the analysis of the hydrogen content must rely entirely on L$\alpha$.

The results for van~Maanen~2 are summarized in Table~\ref{tabres}.
For the abundances, we include three different error sources:
uncertainties from effective temperature, gravity, and from
the choice of the equation of state.
These systematic effects dominate the total error whereas
pure statistical contributions are of minor importance.
Formal statistical errors are comparitively low and do not
reflect the true uncertainties. Therefore, we have determined
the individual contributions from the systematic error sources
using visual comparisons to
models with different parameters and have added the contributions
quadratically.
As mentioned in Sect.~\ref{secatm},
non-ideal effects are not important in the somewhat hotter stars
L\,745-46A and Ross\,640. However, a weak influence can be observed in
van~Maanen~2. If we use the Saha equation instead of our standard EOS, then
the gas pressure increases whereas the electron density decreases and
the absorption lines -- especially the wings -- are in general somewhat
broader. The abundances have to be changed by about 0.1 to 0.2~dex to account
for this effect.

\subsection{LHS\,1126, LP\,658-2, and BPM\,4729}

The analysis of van~Maanen~2 shows the importance of the ultraviolet region
for the temperature determination and the question of the hydrogen abundance.
It seems, therefore, worthwhile to investigate the ultraviolet spectra of
other white dwarfs with similar effective temperature. In this section,
we present HST/FOS observations of LHS\,1126, LP\,658-2, and BPM\,4729,
the coolest white dwarfs in our sample. The observations have been obtained
in guaranteed time but have not been published so far since the fluxes differed
decisively from the predictions of preexisting models raising some doubts about 
the accuracy of the FOS calibration.
However, without going into details, we 
consulted with Ronald Downes of the Space Telescope Science 
Institute and other members of the FOS team and have concluded 
that the data reduced with the standard FOS pipeline are -- to 
the best of our knowledge -- calibrated correctly.

\begin{figure*}[tbp]
\centering
\includegraphics[width=1.0\textwidth]{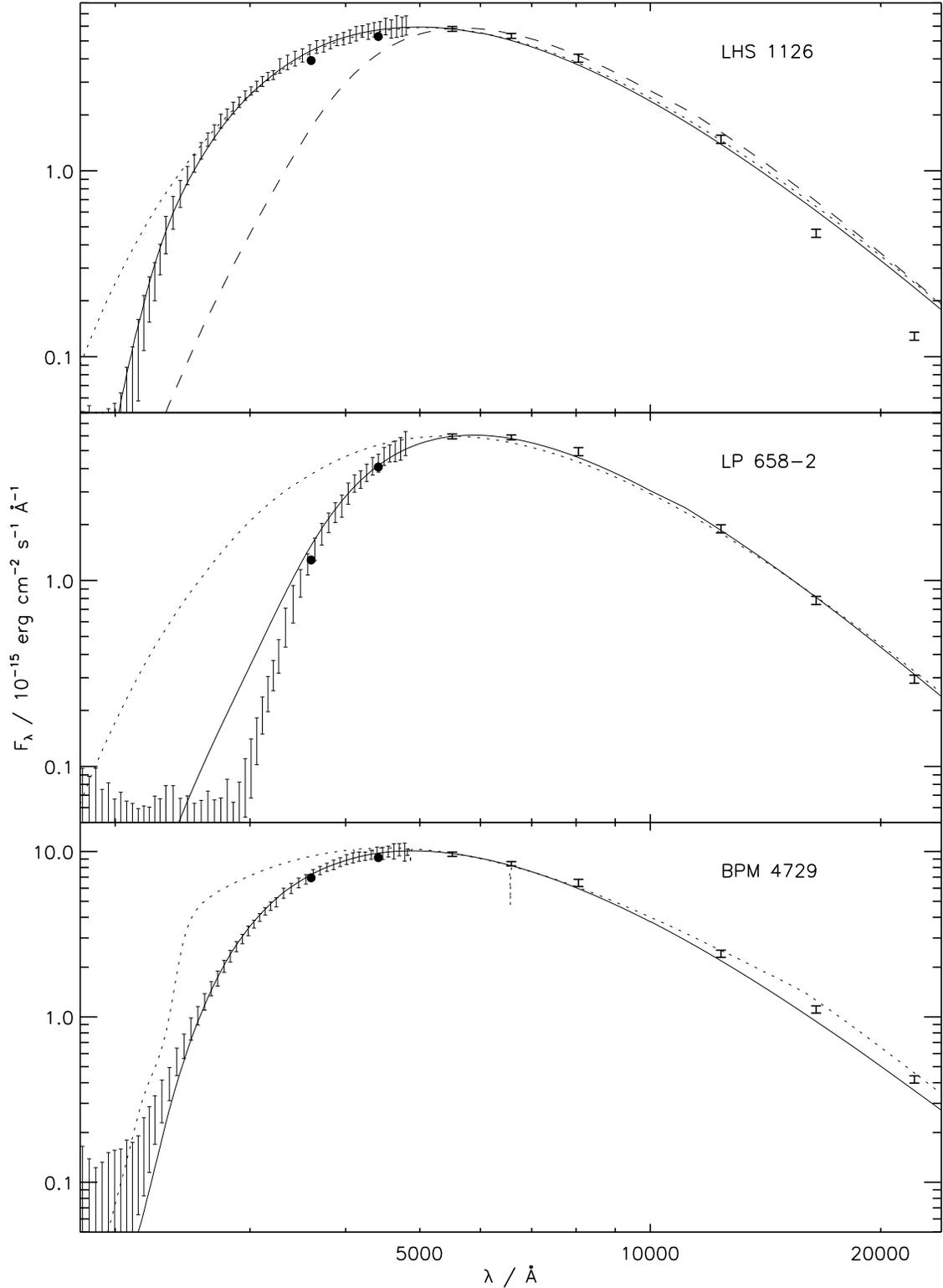}
\caption{HST/FOS spectra (bars) of LHS\,1126 (top), LP\,658-2 (middle), and
BPM\,4729 (bottom) combined with $VRIJHK$ photometry from Bergeron et al.
(\cite{Bergeron01}). Additional photometry in the $B$ (Bergeron et al.
\cite{Bergeron01}) and $U$ band (averages from the compilation of
McCook \& Sion \cite{McCookSion99}) is plotted as filled circles.
The observations are compared to varied model spectra. LHS\,1126:
$T_{\rm eff} = 5400$~K, $\log g = 7.9$, ${\rm H/He}=3\cdot10^{-6}$ (solid line),
pure helium (dotted), and ${\rm H/He}=1\cdot10^{-2}$ (dashed).
LP\,658-2: $T_{\rm eff} = 5050$~K, $\log g = 8.3$, ${\rm H/He}=5\cdot10^{-4}$
(solid), pure helium (dotted). BPM\,4729: $T_{\rm eff} = 5500$~K,
$\log g = 8.21$, ${\rm H/He}=3\cdot10^{-5}$ (solid), $T_{\rm eff} = 5730$~K,
$\log g = 8.21$, pure hydrogen (dotted). All models have been fitted to the
$V$ magnitude}
\label{figcool}
\end{figure*}

LHS\,1126 (WD\,0038$-$226) is one of the few ${\rm C}_2{\rm H}$ stars,
a spectral class introduced by Schmidt et al. (\cite{Schmidt95}). These objects
exhibit broad optical absorption features which have at first been interpreted
as blueshifted ${\rm C}_2$ Swan bands. Schmidt et al., however, showed
that ${\rm C}_2{\rm H}$ is the most probable carbon and hydrogen compound to
be formed under the conditions in LHS\,1126. 
The strong infrared flux deficiency reported by
Wickramasinghe et al. (\cite{Wick82}) could be
explained by Bergeron et al. (\cite{Bergeron94}) using collision-induced
absorption of ${\rm H}_2$ with He.

Bergeron et al. (\cite{Bergeron97}, \cite{Bergeron01}) have determined
$T_{\rm eff} = 5400 \pm 170$~K, $\log g = 7.91 \pm 0.17$, and
$\log {\rm He/H} = 1.86$ from optical/infrared photometry together with the
trigonometric parallax. In Fig.~\ref{figcool} we present a model calculation
(dashed line) using the parameters from Bergeron et al. and compare it
with the HST/FOS spectrum and the photometry. The model flux is significantly
too high at infrared wavelengths. This could be expected since we do not include
collision-induced absorption in our calculations. More severe is, however,
the failure in the ultraviolet. We have included absorption by L$\alpha$ in
the same way as for L\,745-46A and Ross\,640 (see Sect.~\ref{secatm} and
Paper\,I). The high hydrogen abundance of LHS\,1126 results in an extremely
strong absorption.

The observed L$\alpha$ profile can be fitted much better if the abundance
is reduced to ${\rm H/He} = 3 \cdot 10^{-6}$ (solid line in Fig.~\ref{figcool}).
L$\alpha$ absorption is still significant as can be seen from the comparison
with a pure helium atmosphere (dotted line).
The low hydrogen abundance implied by the ultraviolet flux poses a severe
problem since a high value is required to explain the infrared flux deficiency
by collision-induced absorption. Our approach to model the L$\alpha$ wing
has been tested successfully at about 2000~K higher effective temperatures.
It is possible that this approach fails at lower temperatures although
the observed profile can be fitted very well. It could nevertheless be expected
that L$\alpha$ absorption must be strong for an abundance of
${\rm H/He} \approx 10^{-2}$. However, the observations do not show
a strong UV flux deficiency.

A much simpler atmospheric composition is found in LP\,658-2
(WD\,0552$-$041)
which is classified DZ due to the presence of sharp Ca\,II lines
(Eggen \& Greenstein \cite{EG}). Bergeron et al. (\cite{Bergeron01}) have fitted
the optical and ultraviolet photometry using a pure helium atmosphere
with $T_{\rm eff} = 5060 \pm 60$~K and $\log g = 8.31 \pm 0.02$.
In Fig.~\ref{figcool} we compare a model (dotted line) 
with these parameters to the observed flux distribution.
The optical and infrared photometry can be reproduced but the ultraviolet
part is again unsatisfactory. If the flux deficiency is attributed to the
absorption from L$\alpha$ then about ${\rm H/He}=5\cdot 10^{-4}$ would be
necessary (solid line in Fig.~\ref{figcool}). In contrast with LHS\,1126,
the shape of the absorption feature cannot be reproduced in detail. 
Higher hydrogen abundances seem to be necessary for $\lambda < 3000$~\AA\ but
the absorption at longer wavelengths would then be too strong. Our approach for
L$\alpha$ seems to overestimate the absorption in the outermost wing.

The third cool object with FOS observations is BPM\,4729 (WD\,0752$-$676).
Wickramasinghe \& Bessell (\cite{Wick79}) have detected weak H$\alpha$ and
H$\beta$ lines. Bergeron et al. (\cite{Bergeron97}, \cite{Bergeron01})
could fit the optical/infrared photometry and H$\alpha$ with
$T_{\rm eff} = 5730 \pm 110$~K, $\log g = 8.21 \pm 0.09$, and a pure
hydrogen atmosphere. We compare in Fig.~\ref{figcool} a pure hydrogen model
(dotted line) to the observed flux distribution. Again, the ultraviolet part
where L$\alpha$ is important cannot be reproduced. The predicted profile
for L$\alpha$ in a {\em hydrogen} atmosphere is completely at odds
with the observation.
The ultraviolet part looks more like a helium-dominated atmosphere similar to
LHS\,1126 or LP\,658-2. It is indeed possible to fit this region with
${\rm H/He} = 3 \cdot 10^{-5}$ at about $T_{\rm eff} = 5500$~K.
However, optical hydrogen lines are not visible under these conditions
raising doubts on the validity of this solution.

All three objects presented here show problems if the infrared and ultraviolet
parts of the electromagnetic spectrum are to be fitted simultaneously.
In LP\,658-2, additional absorption, most probably from L$\alpha$, seems to be
present which would not affect severely the solution from the
optical/IR photometry. In LHS\,1126 and BPM\,4729, however, the ultraviolet
and infrared observations require completely different atmospheric compositions.
We should note that the HST observations are consistent with existing
photometry in the $B$ and $U$ bands which is plotted as filled circles
in Fig.~\ref{figcool}. Therefore, the discrepancies between the observations
and some models cannot be attributed to remaining uncertainties in
the FOS calibration.

Bergeron et al. (\cite{Bergeron97}, \cite{Bergeron01}) have also noted 
difficulties in fitting the blue energy distribution. They have tried to explain
this with a missing opacity, attributed to a pseudocontinuum originating
from the Lyman edge, for which no physical discription exists. However, one
source of opacity which necessarily is present and which is very strong at
cool temperatures is the L$\alpha$ line. The ultraviolet part of the
spectrum is therefore essential for a determination of the H/He ratio.
We have shown with our new theoretical
discription that at least in He-rich objects the wing extends into the blue
part of the optical spectrum. Under these extreme conditions our line
profile
calculations may be only a first approximation and we cannot yet present 
satisfactory fits. It seems to be important, however, to consider the whole
energy distribution of these objects from the UV to the IR to achieve a
correct interpretation.

\begin{figure*}[tbp]
\centering
\includegraphics[width=\textwidth]{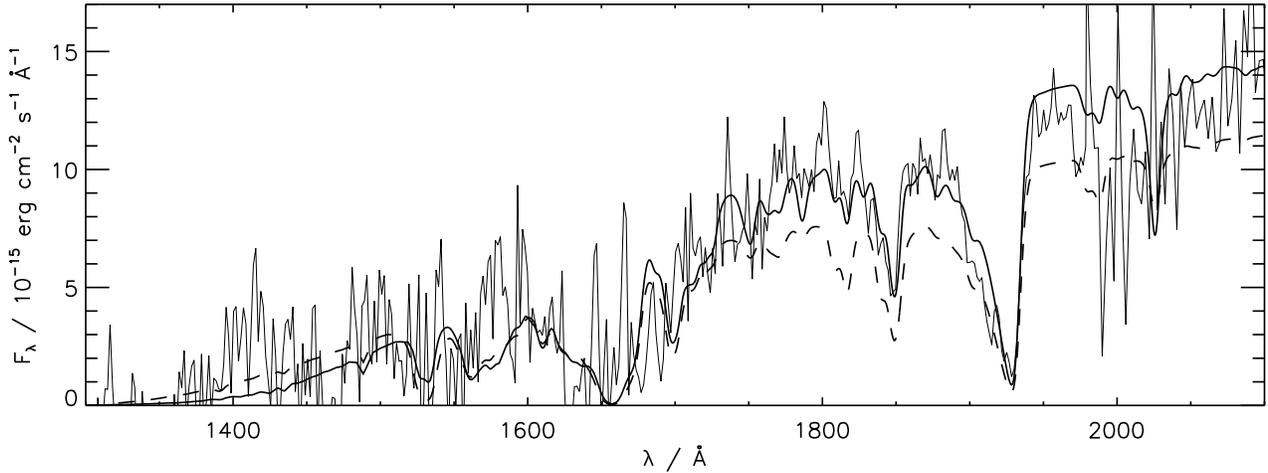}
\caption{L$\alpha$ wing of L\,119-34. The thick continuous curve is a
model with $T_{\rm eff} = 9200$~K and $\log {\rm H/He} = -4.0$.
The dashed curve is a model with $T_{\rm eff} = 8900$~K; the lower temperature
can only partly be compensated by a lower hydrogen abundance
($\log {\rm H/He} = -5.0$). The observed spectrum shows also the C\,I line
at 1930~\AA, Ca\,II at 1840~\AA, and several Fe\,II lines}
\label{figl119}
\end{figure*}

Bergeron et al. (\cite{Bergeron97}, \cite{Bergeron01}) draw far-reaching
conclusions based on their findings of H/He ratios in cool white dwarfs,
e.g. the absence of He atmospheres between 5000 and 6000~K, and their
re-appearance below 5000~K. With our demonstration in this paper that the
parameters from Bergeron et al. fail to explain
the UV flux distribution of several objects it is clear that we have
currently not reached a
consistent interpretation of these objects. The question of H vs. He
atmospheres or the H/He ratio in mixed atmospheres needs further study.

\subsection{L\,119-34 and K\,789-37}

L\,119-34 (WD\,2216$-$657) and K\,789-37 (WD\,1225$-$079) have similar
effective temperatures of about 10\,000~K. IUE spectra are available for both
objects, for K\,789-37 we can also use an optical spectrum (3600--5000~\AA)
obtained with the 1.9m telescope of the South African Astronomical Observatory
(Koester et al. \cite{Koester90}).
K\,789-37 exhibits H$\alpha$, H$\beta$, Mg\,I, and Ca\,II in optical spectra.
The IUE spectrum adds Mg\,II, Si\,I, Fe\,I, and Fe\,II.
We use the IUE spectrum, the visual magnitude
($V=14.79$) and the Str\"omgren colors ($b-y=0.05$, $u-b=0.16$)
from Kilkenny (\cite{Kilkenny86}, \cite{Kilkenny87}) to define the overall
flux distribution.

We start the analysis of K\,789-37 by investigating the effective
temperature and the hydrogen abundance. Since L$\alpha$ determines the
spectral shape at the shortest UV wavelengths both quantities have to be
analyzed simultaneously. However, only a rough estimate of the
hydrogen abundances is necessary for the temperature determination
because L$\alpha$ depends only weakly on the exact value of the abundance.
If we assume $\log g = 8$ we derive $T_{\rm eff} = 10\,500$~K from a fit to
the $V$ magnitude and the IUE spectrum; a visual inspection of H$\beta$ gives
$\log {\rm H/He} = -3.8$. The Str\"omgren colors are also compatible with
these values: $b-y=0.09$ and $u-b=0.16$ for $T_{\rm eff}=10\,500$~K.
They would, however, favor a somewhat higher temperature of about 11\,000~K.
We can determine the surface gravity only very inaccurately with the
existing data. An estimate is possible due to the different sensitivity
of H$\beta$ and L$\alpha$ to changes in $\log g$ and hydrogen abundance.
The effect of a lower or higher gravity on H$\beta$ can be compensated to
some extent by the hydrogen abundance but this leads eventually to an
incompatible flux at the shortest IUE wavelengths. We estimate a possible
range of $\log g = 8.0 \pm 0.5$. The uncertainty of the gravity implies
errors of $\pm$\,300~K for $T_{\rm eff}$ and $\pm$\,0.5 for $\log {\rm H/He}$.

The new effective temperature is about 1000~K higher than the previous results
from Liebert et al. (\cite{Liebert87}; $T_{\rm eff} = 9700$~K) using
multichannel spectrophotometry and
Koester et al. (\cite{Koester90}; $T_{\rm eff} = 9500 \pm 500$~K) using
IUE and optical spectra. The latter could reproduce the optical spectrum
with 10\,000~K whereas the UV flux of the
model was too high with this temperature. We attribute these problems
to the missing L$\alpha$ opacity in their models.

For the analysis of the metal lines, we use $T_{\rm eff} = 10\,500$~K,
$\log g = 8.0$, and $\log {\rm H/He}=-3.8$. The derived abundances are listed
in Table~\ref{tabres}. The errors represent the uncertainties in effective
temperature and gravity. For magnesium, there are additional uncertainties
due to the line profile of the Mg\,II lines at 2796/2803~\AA. The observed
profile cannot be reproduced exactly, a problem which is also known from
the analysis of Ross\,640 (Paper\,I). The broadening of the magnesium
lines is determined by van-der-Waals interaction with neutral helium.
To reproduce the observed profile, we have increased
the broadening parameter for van-der-Waals interaction ($\gamma_6$)
by a factor of five (see also Zeidler-K.T. et al. \cite{Zeidler}).
We added (quadratically) an additional error of
0.5~dex to account for the uncertainty due to the line profile.
The error of the iron abundance represents also the small differences if the
abundances are derived from Fe\,I or Fe\,II, respectively.
The silicon abundance has to be taken with caution because Si\,I cannot be
identified clearly. The same is true
for aluminum ($\log {\rm Al/He} = -7.75 \pm 0.5$, not listed in
Table~\ref{tabres}). From the non-detection of the C\,I line at 1931~\AA,
we determine an upper limit of $\log {\rm C/He}=-7.5$.

Contrary to K\,789-37, hydrogen lines could not be detected in
existing optical spectra of L\,119-34, until now
(see Zeidler-K.T. et al. \cite{Zeidler}).
However, the shorter
wavelength part of the IUE spectrum indicates strongly the presence of
L$\alpha$. We, therefore, determine simultaneously effective temperature
and hydrogen abundance from the overall flux distribution as given by
the IUE spectrum and the visual magnitude ($V=14.43$, Eggen \cite{Eggen}).
This is possible due to the different influence of $T_{\rm eff}$ and L$\alpha$
on the ultraviolet spectrum. The situation is illustrated in
Fig.\,\ref{figl119}. From a visual comparison we derive
$T_{\rm eff} = 9200 \pm 300$~K and $\log {\rm H/He} = -4.0^{+0.5}_{-1.0}$.
Since the surface gravity cannot be determined with existing observations
we have assumed $\log g = 8.0 \pm 0.5$. This is not critical because
the shape of the ultraviolet spectrum depends only weakly on the gravity.
The Str\"omgren colors $b-y$ and $u-b$ (Zeidler-K.T. et al. \cite{Zeidler})
can also be used for a temperature estimate. They give
$T_{\rm eff} = 9200^{+1000}_{-300}$~K, in agreement with the above value.

\begin{figure}[tbp]
\centering
\includegraphics[width=0.5\textwidth]{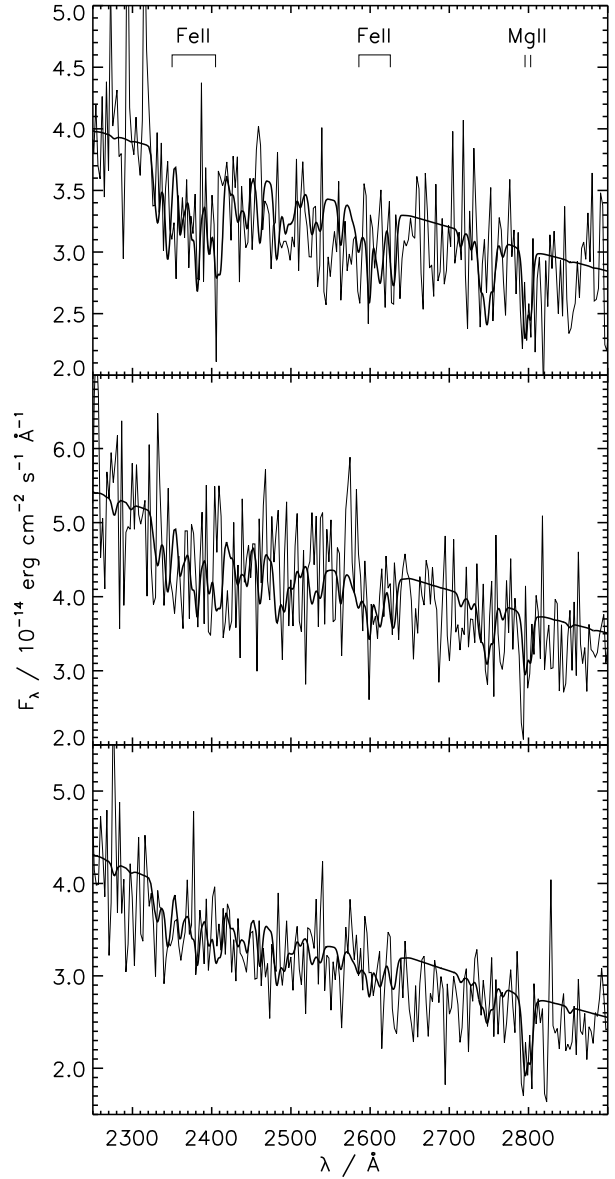}
\caption{IUE spectra of GD\,408 (top), GD\,378 (middle), and GD\,303 (bottom)
compared with model atmospheres for the parameters of Table~\ref{tabres}}
\label{figdbz}
\end{figure}

The element abundances, as determined from the IUE spectrum,
are listed in Table~\ref{tabres}. The errors reflect
the uncertainties in effective temperature and gravity. For silicon, an
additional error of $\pm$\,0.2 is added because of the difficulty to
reproduce all Si\,I lines with the same abundance.
An additional error of $\pm$\,0.25 has also been taken into account for iron
because of abundance differences between Fe\,I and Fe\,II lines.

\begin{table*}[t]
\caption[]{Effective temperatures, gravities, and element abundances
(by number).The errors include systematic uncertainties and are
determined from visual comparisons to model atmospheres. The five hottest
objects are of spectral type DBZ(A) and the others of spectral type DZ(A)}
\begin{flushleft}
\begin{tabular}{l@{}r@{\hspace{0.5em}}llllllll}
\noalign{\smallskip}
\hline
\noalign{\smallskip}
Object         & &$T_{\rm eff}$/K     &$\log g$           &$\log {\rm H/He}$
               &$\log {\rm C/He}$   &$\log {\rm Mg/He}$ &$\log {\rm Si/He}$
               &$\log {\rm Ca/He}$  &$\log {\rm Fe/He}$ \\
\noalign{\smallskip}
\hline
\noalign{\smallskip}
vMa\,2$^a$     &1 &$5700\!\pm\!200$    &$7.9\!\pm\!0.2$    &$-5.0\!\pm\!0.5$
               &                    &$-9.1\!\pm\!0.3$   &$< -9.0$
               &$-10.65\!\pm\!0.2$  &$-9.4\!\pm\!0.2$ \\
L\,745-46A$^b$ &2 &$7500\!\pm\!200$    &$8.0\!\pm\!0.2$    &$-3.1\!\pm\!0.4$
               &$< -8.0$            &$-9.05\!\pm\!0.2$  &$-9.4\!\pm\!0.35$
               &$-10.60\!\pm\!0.2$  &$-9.8\!\pm\!0.2$ \\
Ross\,640$^b$  &3 &$8500\!\pm\!200$    &$8.0\!\pm\!0.2$    &$-3.3\!\pm\!0.3$
               &$< -9.0$            &$-7.25\!\pm\!0.65$ &$-7.5\!\pm\!0.5$
               &$-8.65\!\pm\!0.45$  &$-8.3\!\pm\!0.15$ \\
L\,119-34$^a$  &4 &$9200\!\pm\!300$    &$8.0\!\pm\!0.5$    &$-4.0^{+0.5}_{-1.0}$
               &$-6.0\!\pm\!0.5$    &$-6.8\!\pm\!0.7$   &$-7.9\!\pm\!0.4$
               &$-9.1^d$            &$-7.95\!\pm\!0.7$ \\
K\,789-37$^a$  &5 &$10500\!\pm\!300$   &$8.0\!\pm\!0.5$    &$-3.8\!\pm\!0.5$
               &$<-7.5$             &$-7.6\!\pm\!0.6$   &$-7.5\!\pm\!0.5$
               &$-7.9\!\pm\!0.2$    &$-7.4\!\pm\!0.3$ \\
GD\,408$^a$    &6 &$13750\!\pm\!250$   &$8.0\!\pm\!0.2^e$  &$-6.0\!\pm\!0.3^e$
               &$< -7.0^h$          &$-8.5\!\pm\!0.5$   &$<-8.0$
               &$-9.6\!\pm\!0.3$    &$-7.5^{+0.5}_{-1.0}$ \\
HS\,2253$^c$   &7 &$14700\!\pm\!500$   &$8.0\!\pm\!0.25$   &$-5.0^{+0.7}_{-0.5}$
               &$< -7.5$            &$-5.4\!\pm\!0.2$   &$-5.5\!\pm\!0.7$
               &$-6.1\!\pm\!0.4$    &$-5.1\!\pm\!0.1$ \\
GD\,40$^c$     &8 &$15150\!\pm\!600$   &$8.0\!\pm\!0.25$   &$-5.3\!\pm\!0.5$
               &$-7.2\!\pm\!0.2$    &$-6.0\!\pm\!0.6$   &$-6.5\!\pm\!0.5$
               &$-6.7\!\pm\!0.3$    &$-6.2\!\pm\!0.1$ \\
GD\,378$^a$    &9 &$17000\!\pm\!800$   &$7.9\!\pm\!0.2^f$  &$-4.0^g$
               &$< -6.7^h$          &$-6.5\!\pm\!0.8$   &$<-6.5$
               &$-8.15\!\pm\!0.5$   &$<-5.5$ \\ 
GD\,303$^a$    &10 &$18000\!\pm\!1000$  &$7.8\!\pm\!0.2^f$  &$< -5.5^g$
               &$< -6.5^h$          &$-5.75\!\pm\!0.5$  &$<-7.0$
               &$-7.75\!\pm\!0.5$   &$-5.5^{+0.5}_{-1.5}$ \\
\noalign{\smallskip}
\hline
\noalign{\smallskip}
\multicolumn{9}{l}{$^a$ this paper; $^b$ Paper\,I (Koester \& Wolff \cite{KW});
                   $^c$ Friedrich et al. (\cite{Friedrich});
                   $^d$ Zeidler-K.T. et al. (\cite{Zeidler});}\\
\multicolumn{9}{l}{$^e$ Weidemann \& Koester (\cite{WK});
                   $^f$ Oke et al. (\cite{Oke84});
                   $^g$ Sion et al. (\cite{Sion88});
                   $^h$ Wegner \& Nelan (\cite{WN})}\\
\label{tabres}
\end{tabular}
\end{flushleft}
\end{table*}

\subsection{GD\,408, GD\,378, and GD\,303}

GD\,408 (WD\,0002+729), GD\,378 (WD\,1822+410), and GD\,303 (WD\,1011+570)
are three DBZ stars with effective temperatures well above 10\,000~K.
Photospheric Ca\,II lines could be detected in all three objects but other
metal determinations do not exist in the literature. We examine the IUE
spectra and the optical observations of Sion et al. (\cite{Sion88}).

Weidemann \& Koester (\cite{WK}) have determined $T_{\rm eff}$ from
multichannel spectrophotometry and $\log g$ with the help of the trigonometric
parallax for GD\,408. Oke et al. (\cite{Oke84}) used multichannel data
to determine $T_{\rm eff}$ and $\log g$ for GD\,378 and GD\,303. We have used
the results for the gravity in our analysis but have tested the previous
temperature determinations using the IUE spectra and the visual magnitudes.
Our results for GD\,408 and GD\,303 (see Table~\ref{tabres}) are compatible
with the previous values; for GD\,378 we determine a temperature
about 1500~K hotter than Oke et al. (\cite{Oke84}).

The IUE spectra of all three stars reveal weak lines from Mg\,II and Fe\,II
(Fig.~\ref{figdbz}). Since these features are present in all
three spectra we are confident that they are not observational
artifacts. The derived abundances are listed in Table~\ref{tabres}.
The errors include uncertainties due to errors in effective temperature
and gravity.
We have also analyzed the optical spectra of Sion et al. (\cite{Sion88}) again.
Our new values for the calcium abundance
are about two orders of magnitude higher than the values
derived by Dupuis et al. (\cite{Dupuis93b}) because the scaling of the
equivalent width by the square root of the abundance, as used by Dupuis et al.
in their analysis, does not give reliable results here.

\section{Discussion}

We have analyzed effective temperatures, gravities, and surface abundances
for six DZ/DBZ white dwarfs. For the following discussion,
we add the results from Paper\,I and the two DBZ white dwarfs from
Friedrich et al. (\cite{Friedrich}). These ten stars span virtually the 
complete range of temperatures of DZ/DBZ white dwarfs. They are the only
objects with ultraviolet observations which provide the best possibility to
investigate the whole spectrum of metals; in optical observations,
only Ca\,II is found in general.
We collect the parameters for all stars in Table~\ref{tabres}.

\begin{figure*}[tbp]
\centering
\includegraphics[width=1.0\textwidth]{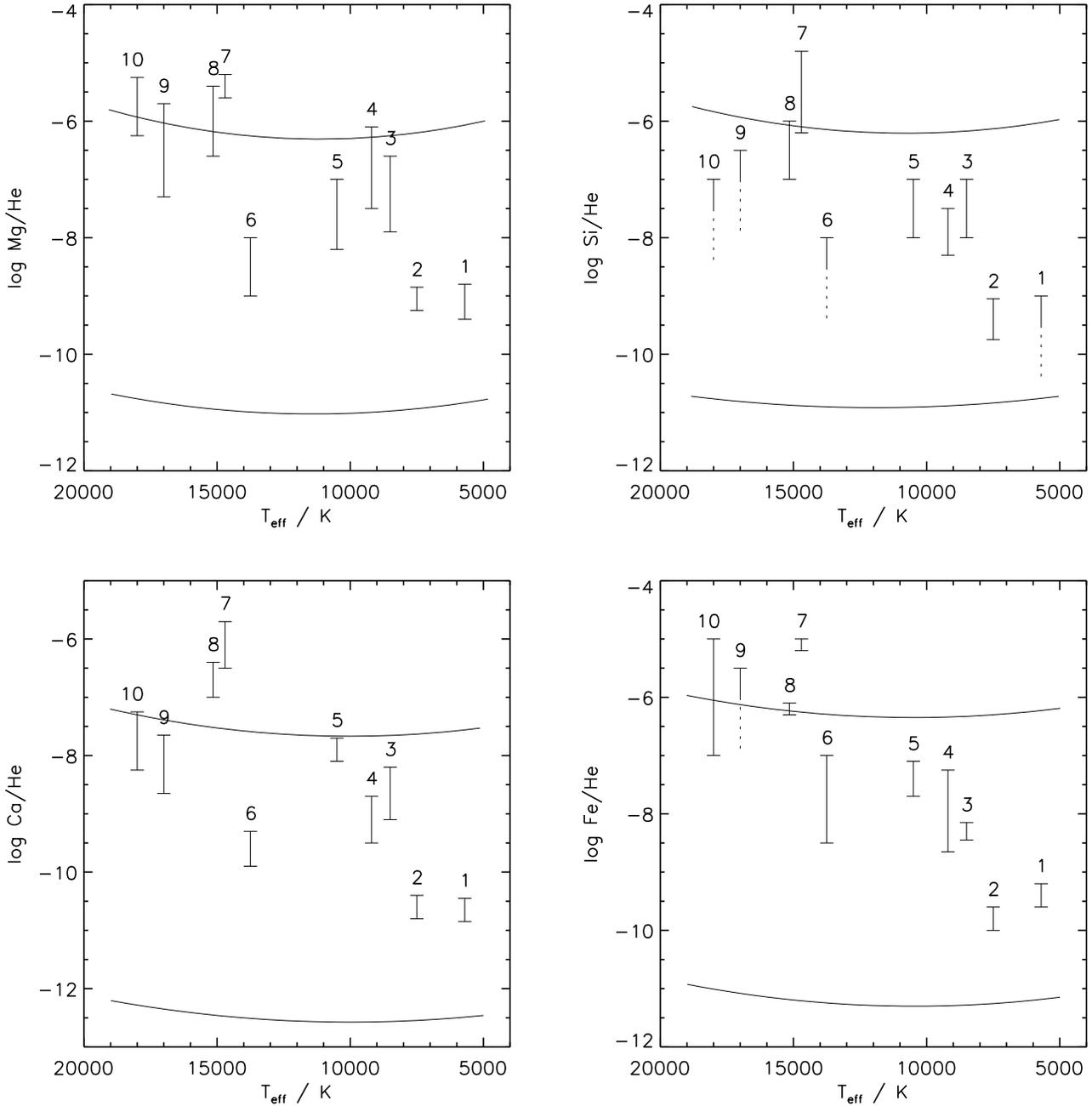}
\caption{Observed metal abundances (error bars) compared with the predicted
ranges from the two-phase accretion model of Dupuis et al. (\cite{Dupuis93b}).
The lower continuous curve is the equilibrium abundance reached in the
low accretion phase. The upper curve is the maximum abundance
during an encounter with an interstellar cloud.
Observational upper limits are denoted by dashed bars. The labels correspond
to the numbering in Table~\ref{tabres}}
\label{figabun}
\end{figure*}

We start with a comparison of the observed abundances with the predicted ranges
from the model of Dupuis et al. (\cite{Dupuis93b}) in Fig.~\ref{figabun}.
The two continuous curves in each panel represent the lowest and highest
possible abundances at each effective temperature. The lower curves correspond
to the equilibrium abundances reached in the low accretion phase
($\dot{M}\approx 5\cdot10^{-20}\,M_{\odot}\,{\rm yr}^{-1}$
for 49 million years in the model of Dupuis et al.) whereas the
upper curves correspond to the maximum abundance reached during passages
through interstellar clouds
($\dot{M}\approx 5\cdot10^{-15}\,M_{\odot}\,{\rm yr}^{-1}$
for one million years).
Both curves should be taken with some caution since the actual values of
the accretion rates are not known exactly. Moreover, the upper curve
depends also partly on the duration of the cloud encounter since an
equilibrium between accretion and diffusion is not reached at all temperatures
during the assumed passage value of $10^6$ years.

The comparison shows that the model assumptions can account in general
for the observed metal abundances since most observed values
are well between the two limits of the model predictions.
Only the observations of GD\,40 and HS\,2253+8023 
(numbers 7 and 8 in Fig.~\ref{figabun}) are in part
above the upper limits but this can be attributed to the mentioned
uncertainties.
The data in Fig.~\ref{figabun} seem also to indicate a correlation of the
abundances with effective temperature. This is, however, a selection effect
since at higher temperatures only high metal abundances can be detected
in the low resolution and low signal-to-noise observations normally used to
identify DZ white dwarfs (see e.g. Zeidler-K.T. et al. \cite{Zeidler} for
quantitative estimates of the visibility ranges).

\begin{figure*}[tbp]
\centering
\includegraphics[width=1.0\textwidth]{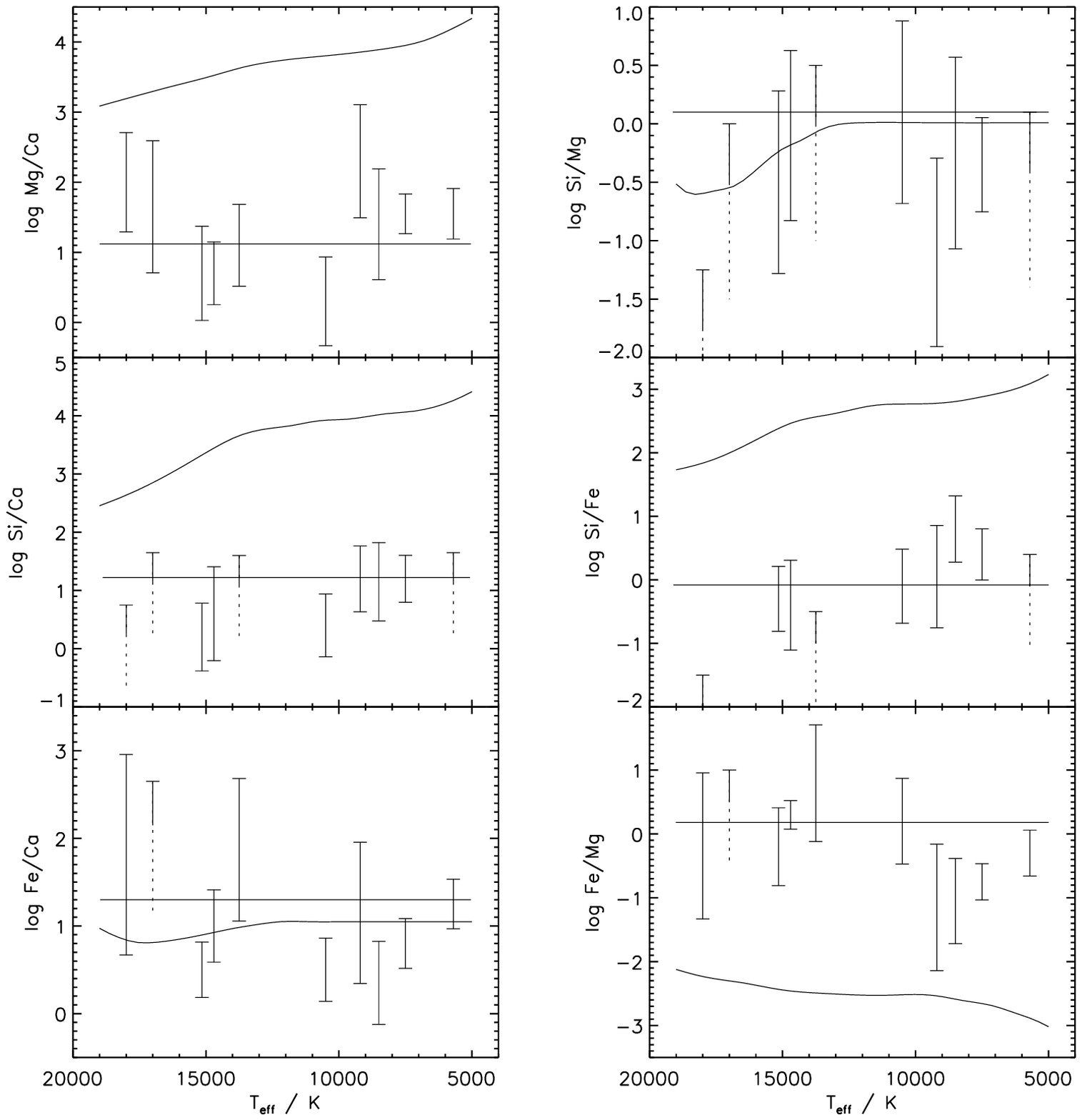}
\caption{Observed metal abundance ratios (error bars) compared with the
predicted ratios from Dupuis et al. (\cite{Dupuis93b}). The two continuous
curves give the possible range if the accreted material has solar composition.
Upper limits from the observations are denoted by dashed bars}
\label{figratio}
\end{figure*}

Further information about the accretion process can be extracted from
the ratios of the metal abundances. Fig.~\ref{figratio} shows the values for
the six independent combinations of the four observed metals.
Again, the continuous curves define
the allowed ranges from the model of Dupuis et al. (\cite{Dupuis93b}).
The straight line is the solar abundance ratio which is reached in the
first moments of a cloud encounter if the accreted material has solar
composition. Afterwards, the ratio is altered due to diffusion and approaches
the equilibrium value (not shown in Fig.~\ref{figratio}) which is always
within 0.5~dex of the solar ratio.
The second curve in Fig.~\ref{figratio} corresponds to the maximum
(or minimum) value reached shortly after a cloud encounter when the abundance of
the element with the shorter diffusion time scale is again near the value
during low accretion. Afterwards, the ratio turns back to the equilibrium value.
At any time, the observed abundance ratio must be
between the two curves shown in Fig.~\ref{figratio}.
The difference between them depends on the
diffusion time scales of the two elements.

Most individual abundance ratios fall into the predicted range. They are also
often near the solar value (and the equlibrium value) which can be seen well
in the Mg/Ca and Fe/Mg diagrams. This implies that we observe the
DZ stars mostly during the high accretion phase -- in agreement with the
observation that the abundances in Fig.~\ref{figabun} are near the upper curve.
The deviations from the predicted regions
could be attributed to individual variations of the chemical
composition of the accreted material which could be different from star to star.
However, we follow here the discussion of Dupuis et al. (\cite{Dupuis93b}) who 
explained these differences by the same changes of the composition for all
stars. This conclusion is supported by systematic
tendencies which were not clearly present in previous studies:
Several observed values especially of the Si/Ca and Si/Fe ratios
are below the allowed range.

\begin{figure}[tbp]
\centering
\includegraphics[width=0.5\textwidth]{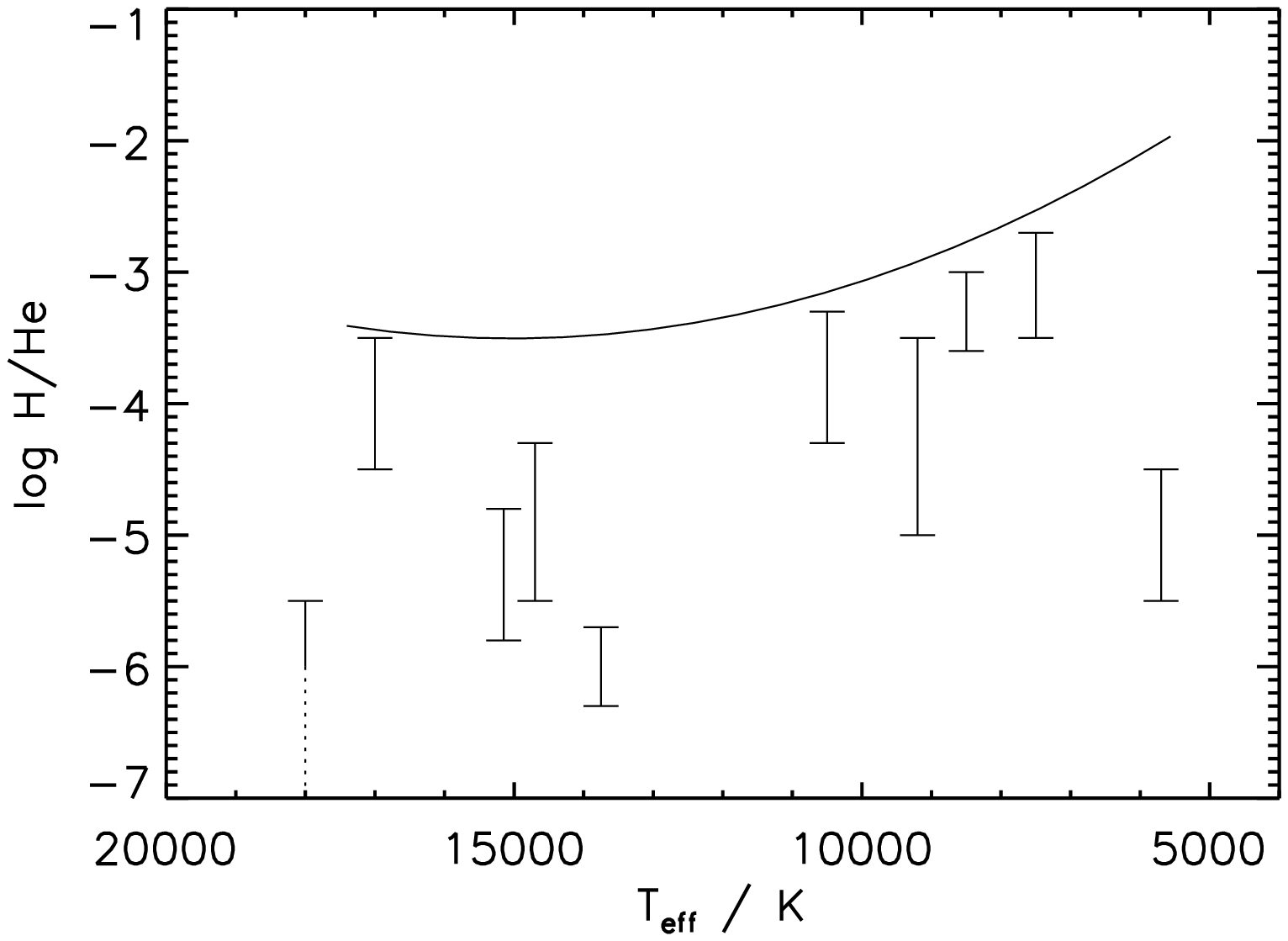}
\caption{Observed hydrogen abundances compared to the expected abundance
for continuous accretion of material with solar composition
with $\dot{M}=10^{-18}~M_{\odot}\,{\rm yr}^{-1}$.}
\label{fighyd1}
\end{figure}

If the abundance of one element in the accreted material is changed then the
whole region of possible ratios in Fig.~\ref{figratio} is
shifted by that amount. If we want virtually all measured values to fall
into the predicted range then the silicon abundance must be changed
by a factor of $\approx$\,0.2 ($-$0.7~dex) and the calcium abundance by a factor
of $\approx$\,2 (0.3~dex). Similar values were also determined by
Dupuis et al. (\cite{Dupuis93b}).

The underabundance of silicon is interesting in connection with the assumption
that metals are accreted onto white dwarfs in the form of grains.
Silicon has the lowest
condensation temperature of the four metals usually found in DZ spectra
(see e.g. Spitzer \cite{Spitzer}, Fig.~9.1). During the formation of grains
for example  in the winds of cool giants it condenses later than the other
elements.
It has, therefore, a lower depletion factor in the gas phase of the interstellar
medium and may be underabundant in the grains accreted onto white dwarfs.
During the infall onto the white dwarf silicon could be evaporated earlier than
other metals and may partly not reach the surface. Other elements with higher
condensation temperatures would survive the accretion process better.
Note that calcium, which is probably overabundant, has the highest
condensation temperature of the four metals.

In contrast to heavier elements, hydrogen is not accreted in the form of
grains. As mentioned in the introduction, it can only be observed
in DZ white dwarfs at unexpectedly low abundances.
Nine of our objects exhibit hydrogen, in two of them we could show for the
first time its presence through absorption by L$\alpha$.
The distribution of hydrogen abundances against the effective temperatures
of the stars is shown in
Fig.~\ref{fighyd1}. As in previous investigations, a correlation with
effective temperature cannot be observed. We show as a comparison also
the expected hydrogen abundance for continuous accretion of
$\dot{M}=10^{-18}~M_{\odot}\,{\rm yr}^{-1}$ (Dupuis et al. \cite{Dupuis93b}).
The calculation assumes solar composition of the accreted material and
a start of the accretion at $T_{\rm eff} = 20\,000$~K. The exact choice of the
starting point is not critical because of the
rapidly increasing cooling ages towards lower temperatures. Fig.~\ref{fighyd1}
shows that the actual average accretion rate of hydrogen must be lower
than $\dot{M}=10^{-18}~M_{\odot}\,{\rm yr}^{-1}$ for our sample objects.
On the other hand, the
accretion rate for metals -- averaged over $5\cdot10^7$ years -- in the model
of Dupuis et al. is about $\dot{M}=10^{-16}~M_{\odot}\,{\rm yr}^{-1}$.
Therefore, the hydrogen accretion rate is at least two orders of magnitude
lower than the metal accretion rate, even for those objects with the highest
hydrogen abundances.

\begin{figure}[tbp]
\centering
\includegraphics[width=0.5\textwidth]{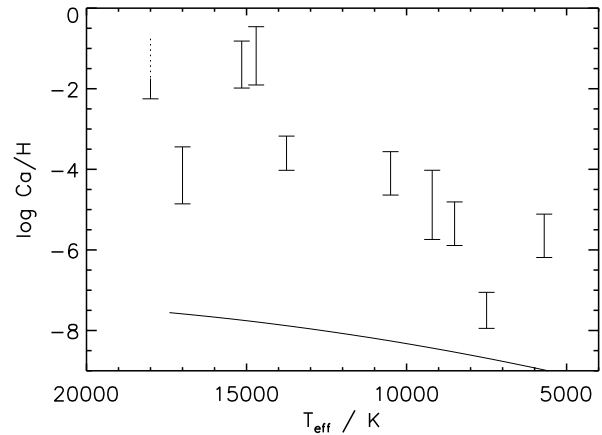}
\caption{Calcium-to-hydrogen ratios compared to the maximum Ca/H ratio
which could be expected for accretion of material with solar composition.}
\label{fighyd2}
\end{figure}

In this context, the ratio of hydrogen to other metals is also interesting.
Fig.~\ref{fighyd2} shows the observed calcium-to-hydrogen ratios.
The weak tendency in temperature which might be observed
in the data is due to a similar effect visible in the calcium abundances
(see above). Fig.~\ref{fighyd2} shows also the maximum value of {\rm Ca/H}
which can be expected for accretion of material with solar composition
within the model of Dupuis et al. (\cite{Dupuis93b}). All observed values
are significantly higher than the theoretical upper limit. This shows again
that hydrogen is not accreted at solar abundance ratios relative to metals.

\section{Conclusions}

The element abundances of the ten DZ and DBZ white dwarfs presented in this
paper form a homogeneous set of data which has been compared to the
two-phase accretion model of Dupuis et al. (\cite{Dupuis93b}). The
absolute values are in the expected range for accretion from the interstellar
medium. Most metal abundance ratios are also compatible with the predictions.
Small deviations could be an indication of non-solar metal-to-metal ratios in
the accreted material. This tendency is more clearly visible in our improved
data set than in the previous investigation of Dupuis et al. (\cite{Dupuis93b}).
Hydrogen is present in virtually all of our objects. However, the average
accretion rate of hydrogen must be more than two orders of magnitude lower
than the average accretion rate of the metals -- even in those objects with
the highest hydrogen abundances. Therefore, the fate of hydrogen remains
unclear.

Our success in modeling L$\alpha$ absorption in L\,745-46A and Ross\,640
(Paper\,I) provides a new means of detecting hydrogen in those cases
where
the Balmer lines are already very weak. This method has been used
to determine the hydrogen abundance in L\,119-34 and van~Maanen~2. This
is important especially for the latter object because H$\alpha$ is not
visible at this combination of effective temperature and hydrogen
abundance. In three other white dwarfs with temperatures similar to
van~Maanen~2 the ultraviolet region of the spectrum
indicates a completely different chemical
composition than that implied by the optical and infrared
photometry. However, we are at the moment not in the position to solve
this discrepancy.
The analysis of van~Maanen~2 shows also that the analysis of optical and
infrared photometry can lead to wrong effective temperatures if the
absorption by metals in the ultraviolet is not considered. The temperature
would be about 1000~K higher without appropriate blanketing in the UV.

\begin{acknowledgements}
We are grateful to Jay~Holberg (Tucson) who additionally processed
some of the IUE spectra used in this paper and to the referee
Martin~Barstow (Leicester) for valuable comments.
JL acknowledges the contributions
of Roger Angel and Richard Allen (Steward Observatory) to obtain the
FOS data and the help of Ronald Downes (STScI) to ascertain that the data
reduced with the standard FOS pipeline is -- to the best of our knowledge --
calibrated correctly.
This work has been supported by the Deutsches Zentrum f\"ur Luft- und
Raumfahrt (DLR) under grant 50 OR 96173. We have made use of the Simbad
data base, operated at CDS, Strasbourg, France.
\end{acknowledgements}

\listofobjects


\begin{thebibliography}{}

\bibitem[1991]{AKi91} Allard, N., \& Kielkopf, J. 1991, A\&A, 242, 133
\bibitem[1994]{Bergeron94} Bergeron, P., Ruiz, M.-T., Leggett, S.K.,
        Saumon, D., \& Wesemael, F. 1994, ApJ, 423, 456
\bibitem[1997]{Bergeron97} Bergeron, P., Ruiz, M.T., \& Leggett, S.K. 1997,
        ApJS, 108, 339
\bibitem[2001]{Bergeron01} Bergeron, P., Legget, S.K., \& Ruiz, M.T. 2001,
        ApJS, 133, 413
\bibitem[1992]{Dupuis92} Dupuis, J., Fontaine, G., Pelletier, C., \&
        Wesemael, F. 1992, ApJS, 82, 505
\bibitem[1993a]{Dupuis93a} Dupuis, J., Fontaine, G., Pelletier, C., \&
        Wesemael, F. 1993a, ApJS, 84, 73
\bibitem[1993b]{Dupuis93b} Dupuis, J., Fontaine, G., \& Wesemael, F. 1993b,
        ApJS, 87, 345
\bibitem[1969]{Eggen} Eggen, O.J 1969, ApJ, 157, 287
\bibitem[1965]{EG} Eggen, O.J., \& Greenstein, J.L. 1965, ApJ 141, 83
\bibitem[1997]{ESA} ESA 1997, The Hipparcos and Tycho Catalogues, Vol. 1-16
        (ESA Publications Division, Noordwijk)
\bibitem[1997]{FKB} Finley, D.S., Koester, D., \& Basri, G. 1997, ApJ, 488, 375
\bibitem[1999]{Friedrich} Friedrich, S., Koester, D., Heber, U., Jeffery, C.S.,
        \& Reimers, D. 1999, A\&A, 350, 865
\bibitem[1974]{Grenfell74} Grenfell, T.C. 1974, A\&A 31, 303
\bibitem[1986]{Kilkenny86} Kilkenny, D. 1986, Observatory, 106, 201
\bibitem[1987]{Kilkenny87} Kilkenny, D. 1987, MNRAS, 229, 345
\bibitem[2000]{KW} Koester, D., \& Wolff, B. 2000, A\&A 357, 587 (Paper\,I)
\bibitem[1990]{Koester90} Koester, D., Wegner, G., \& Kilkenny, D. 1990,
        ApJ, 350, 329
\bibitem[1987]{Liebert87} Liebert, J., Wehrse, R., \& Green, R.F. 1987,
        A\&A, 175, 173
\bibitem[1999]{McCookSion99} McCook, G.P., \& Sion, E.M. 1999, ApJS, 121, 1
\bibitem[1990]{Mihalas90} Mihalas, D., Hummer, D.G., Mihalas, B.W., \&
        D\"appen W. 1990, ApJ, 350, 300
\bibitem[1994]{Nichols94} Nichols, J.N., Garhart, M.P., de la Pe\~{n}a, M.D.,
        Levay, K.R. 1994, NASA IUE Newsletter 53
\bibitem[1984]{Oke84} Oke, J.B., Weidemann, V., \& Koester, D. 1984,
        ApJ, 281, 276
\bibitem[2002]{Provencal} Provencal, J.L., Shipman, H.L., Koester, D., 
        Wesemael, F., Bergeron, P. 2002, ApJ, in press
\bibitem[1994]{SC94} Saumon, D., \& Chabrier, G. 1994, Phys.Rev.A, 46, 2084
\bibitem[1995]{SCVH95} Saumon, D., Chabrier, G., \& Van Horn, H.M. 1995, ApJS,
        99, 713
\bibitem[1995]{Schmidt95} Schmidt, G.D., Bergeron, P., \& Fegley, B., Jr. 1995,
        ApJ, 443, 274
\bibitem[1988]{Sion88} Sion, E.M., Aannestad, P.A., \& Kenyon, S.J. 1988,
        ApJ, 330, L55
\bibitem[1978]{Spitzer} Spitzer, L. 1978,
        Physical Processes in the Interstellar Medium, John Wiley \& Sons,
        New York
\bibitem[1972]{Wegner72} Wegner, G. 1972, ApJ 172, 451
\bibitem[1987]{WN} Wegner, G., \& Nelan, E.P. 1987, ApJ, 319, 916
\bibitem[1960]{Weidemann60} Weidemann, V. 1960, ApJ, 131, 638
\bibitem[1991]{WK} Weidemann, V., \& Koester, D. 1991, A\&A, 249, 389
\bibitem[1979]{Wick79} Wickramasinghe, D.T., \& Bessell M.S. 1979, MNRAS 186,
        399
\bibitem[1982]{Wick82} Wickramasinghe, D.T., Allen, D.A., \& Bessell M.S. 1982,
        MNRAS, 198, 473
\bibitem[1994]{Wood94} Wood, M.A. 1994,
        in: The Equation of State in Astrophysics, IAU Coll. 147,
        ed. G. Chabrier, \& E. Schatzman E., Cambridge University Press, 612
\bibitem[1986]{Zeidler} Zeidler-K.T., E.-M., Weidemann, V., \& Koester, D.
        1986, A\&A, 155, 356
\end{thebibliography}
\end{document}